\documentclass[epj,nopacs]{svjour}
\usepackage{graphicx}
\begin{document}
\title{Static meson correlators in 2+1 flavor QCD at non-zero temperature}
\author{A. Bazavov\inst{1} \and P. Petreczky\inst{1}
}                     
%
%
\institute{Physics Department, Brookhaven National Laboratory, Upton, NY 11973, USA}
\date{Received: date / Revised version: date}
%
\abstract{
We study correlation functions of various static meson operators
of size $r$ at non-zero temperature in 2+1 flavor QCD, including
Coulomb gauge fixed operators and Wilson loops with smeared spatial
parts. The numerical calculations are performed on $24^3 \times 6$
lattices using highly improved staggered quark action. 
We discuss possible implications of our findings on 
the temperature dependence of the static energy of $Q\bar Q$ pair.
%
} 
\maketitle
\section{Introduction}
\label{intro}

At high temperature strongly interacting matter undergoes a transition to
a new state called quark gluon plasma (QGP). Creating and exploring such state
of matter is a subject of a large experimental program, see e.g. Ref. \cite{Muller:2006ee}.
At the same time properties of strongly interacting matter at high temperatures
can be studied in ab-initio calculations that rely on lattice regularization  of QCD
(see Ref. \cite{Philipsen:2012nu,Petreczky:2012rq} for recent reviews). 

The suppression of quarkonium production has been suggested by Matsui and
Satz as probe of QGP formation in heavy ion collisions \cite{Matsui:1986dk}.
Interaction between heavy quark and anti-quark is 
expected to be modified by the deconfined medium at high
temperature eventually leading to the dissociation of 
the quarkonium states. The suppression of the quarkonium yields in
heavy ion collisions was indeed observed experimentally (see Ref. \cite{GranierdeCassagnac:2008ke}).
The interpretation of the experimental findings, however, requires the knowledge of quarkonium 
properties at high temperatures among other things (see Refs. \cite{Mocsy:2013syh,Brambilla:2010cs,Rapp:2008tf} 
for recent reviews).
Therefore there has been a significant
effort to study in-medium properties of heavy quarkonium in recent years including
lattice QCD studies. In-medium quarkonium pro\-perties are encoded in the spectral functions.
A commonly used approach to obtain quarkonium spectral functions 
relies on lattice QCD calculation of the correlation 
functions in Euclidean time direction and extraction
of the quarkonium spectral functions using the Maximum Entropy Method 
\cite{Umeda:2002vr,Asakawa:2003re,Datta:2003ww,Jakovac:2006sf,Aarts:2007pk,Ding:2012sp}.
However, it turns out that temporal quarkonium correlators
are remarkably insensitive to the in-medium modification
of the corresponding spectral functions due to the limited
extent of the imaginary time direction $\tau<1/(2 T)$ \cite{Petreczky:2008px}.
The dissolution of the bound state peaks in the spectral functions is
compensated by large enhancement of the spectral function in the threshold region in such a way
that corresponding Euclidean correlation functions do not change significantly 
for $\tau<1/(2T)$ \cite{Mocsy:2007yj}.
This picture is corroborated by the study of the spatial charmonium correlation functions 
\cite{Karsch:2012na}. Unlike temporal correlators spatial correlators can be
calculated for separation larger than $1/(2T)$ and show strong temperature
dependence consistent with dissolution of the bound states \cite{Karsch:2012na}.

Because of the above problems the calculation of the 
quarkonium spectral functions from finite temperature Euclidean time 
correlation functions is very difficult.
An alternative approach is to calculate quarkonium spectral functions
relies on effective theories. Quarkonium is characterized by
three different energy scales: the heavy quark mass $m$, the
inverse quarkonium size $1/r \sim m v$ and the quarkonium binding
energy $m v^2$. Integrating out the largest energy scale $\sim m$ leads to
an effective theory called non-relativistic QCD (NRQCD). This effective theory was combined
with lattice QCD method to study quarkonium properties both at zero
\cite{Dowdall:2011wh} and non-zero temperatures \cite{Aarts:2010ek,Aarts:2011sm}.
Integrating out the scale $m v$ gives an effective theory called 
potential non-relativistic QCD (pNRQCD) \cite{Brambilla:1999xf,Brambilla:2004jw}. 
The Lagrangian of pNRQCD is formulated in terms of singlet and
octet fields of the heavy quark anti-quark pair and the 
heavy quark potential enters as the parameter of the effective
Lagrangian. At non-zero temperature additional scales $T$ and $gT$
appear. Depending on how these scales are related to the above
energy scales the heavy quark potential may be modified by the
medium accordingly \cite{Brambilla:2008cx}. 
At non-zero temperature the potentials also have an imaginary part \cite{Brambilla:2008cx,Laine:2006ns}.
The above effective field theory approach is based on the weak
coupling, namely on the assumption that $m v,~T,~g T \gg \Lambda_{QCD}$.
Within this framework in-medium quarkonium properties can be calculated \cite{Brambilla:2010vq}
and one can also take the static limit and calculate the energy of a static $Q \bar Q$ pair which 
like the potential also has an imaginary part.

It is not clear if in the interesting temperature range the above scale separation holds.
However, we may expect that the binding energy of quarkonium
states is reduced with increasing temperature. Close to the dissolution point
the binding energy is the smallest scale in the problem and all other scales,
including $\Lambda_{QCD}$ can be integrated out. In that regime the potential
of the corresponding effective theory is equal to the energy of $Q \bar Q$ 
pair \cite{Petreczky:2010tk}. Therefore calculation of the static energy
at non-zero temperature in the non-perturbative region 
could be very helpful for the determination of in-medium quarkonium properties.

At zero temperature the energy of static $Q\bar Q$ pair is calculated
by studying correlation function of static meson operators
at large Euclidean time separations. This is not possible
and non-zero temperature due to the limited temporal extent $\tau \le 1/T$.
It has been suggested to extract the static quark potential
at non-zero temperature from the spectral representation of
the simplest finite temperature static meson correlators, namely rectangular Wilson loops and MEM 
\cite{Rothkopf:2009pk,Rothkopf:2011db}. The analysis was performed in quenched approximation.
In this paper
we are going to analyze the temperature dependence of correlation functions of various 
static meson operators in 2+1 flavor QCD across the chiral crossover region
and discuss its implication for the temperature dependence
of the static $Q\bar Q$ energy. Apart from quantitative determination of the static
energy  at non-zero temperature the study of static meson correlators
is interesting because they are simpler than quarkonium correlators
and can provide some insight into the inter-quark interaction in the medium
and the weakly interacting nature of QGP.
One of the aims of the present study is to find out to what extent the temperature
dependence is influenced by the choice of the interpolating meson operator.
Obviously only the features that are not sensitive to the specific choice of
the meson operators are of some relevance. Some preliminary results have been reported
in conference proceedings \cite{Bazavov:2012bq,Bazavov:2012fk}.
\begin{figure*}
\begin{center}
\includegraphics[width=7.7cm]{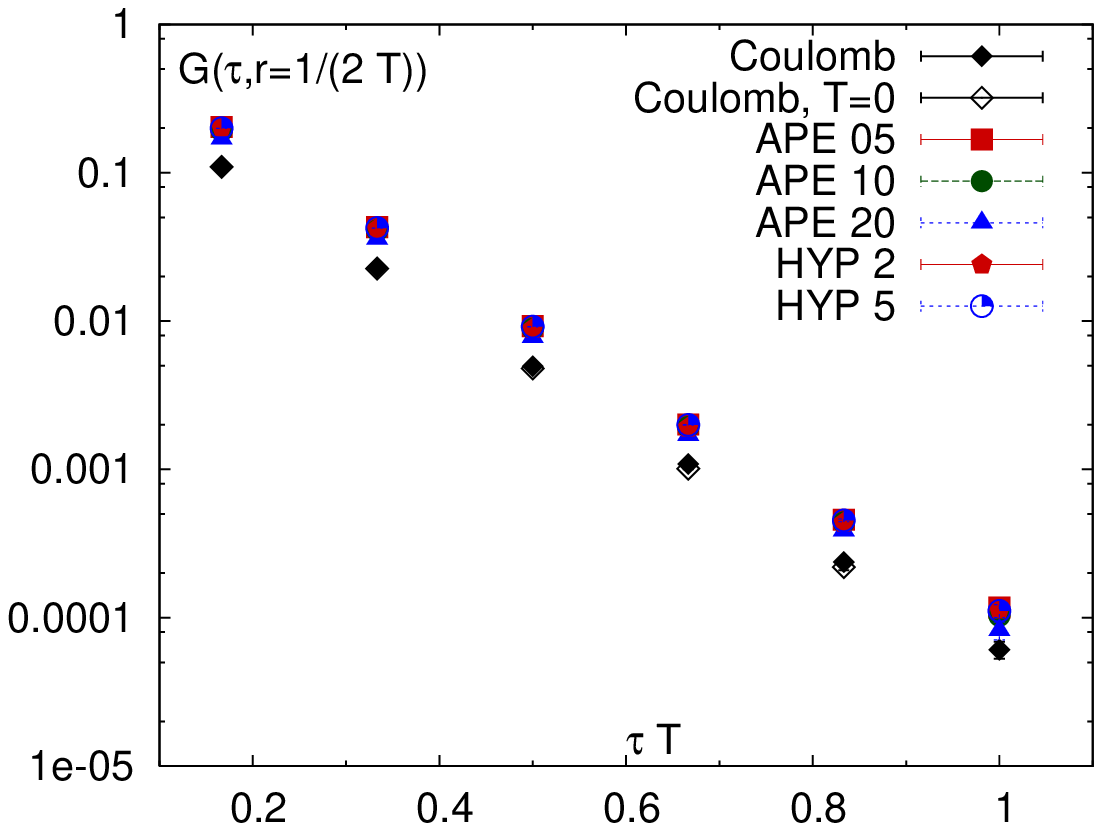}
\includegraphics[width=7.7cm]{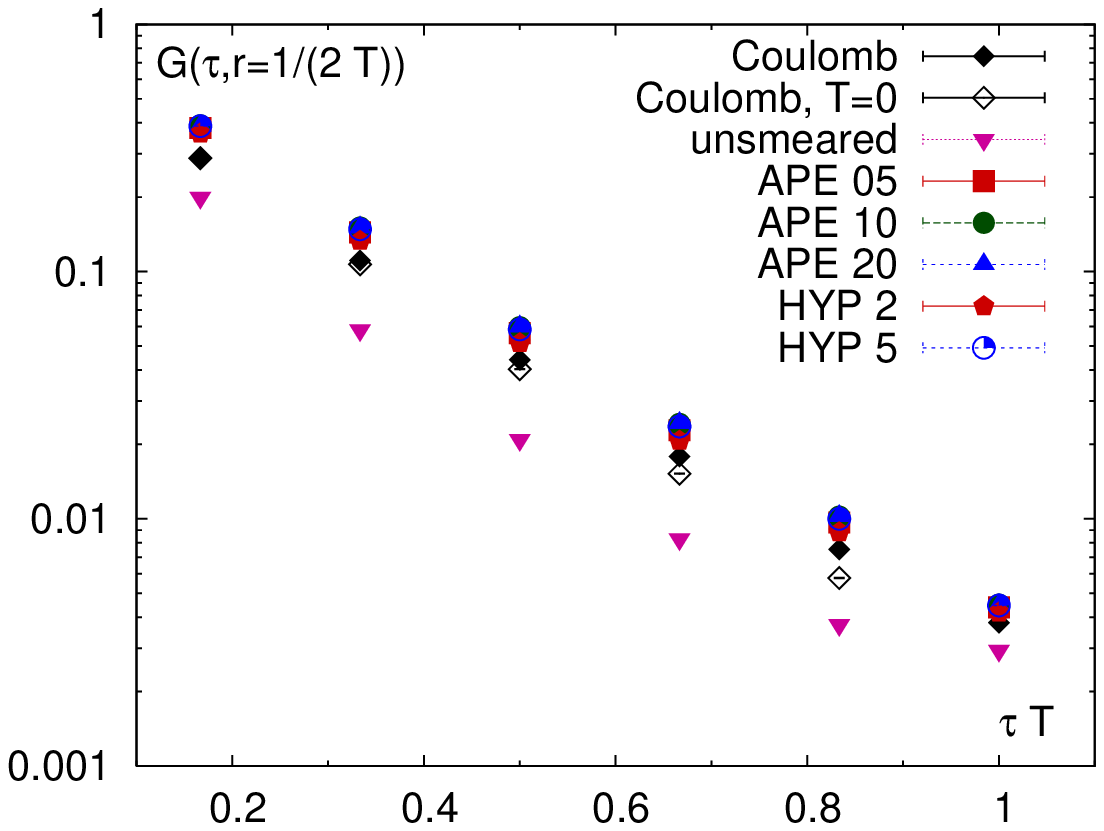}
\end{center}
\begin{center}
\includegraphics[width=7.7cm]{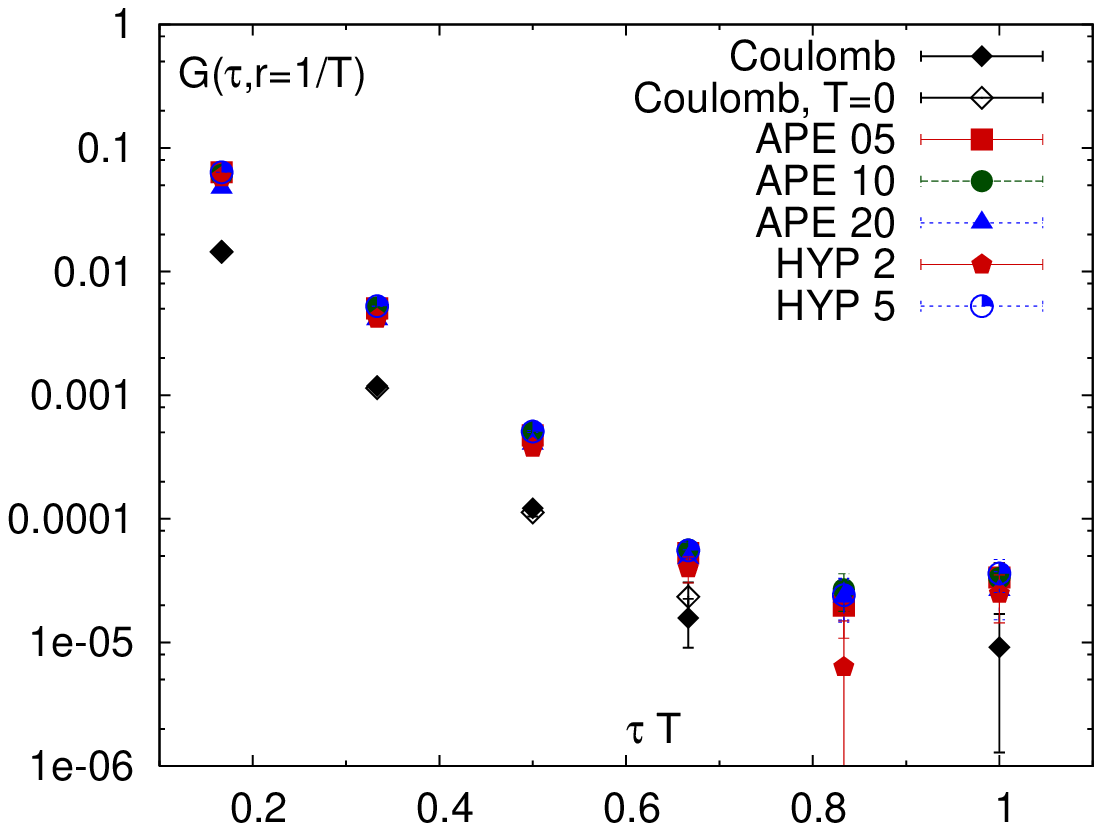}
\includegraphics[width=7.7cm]{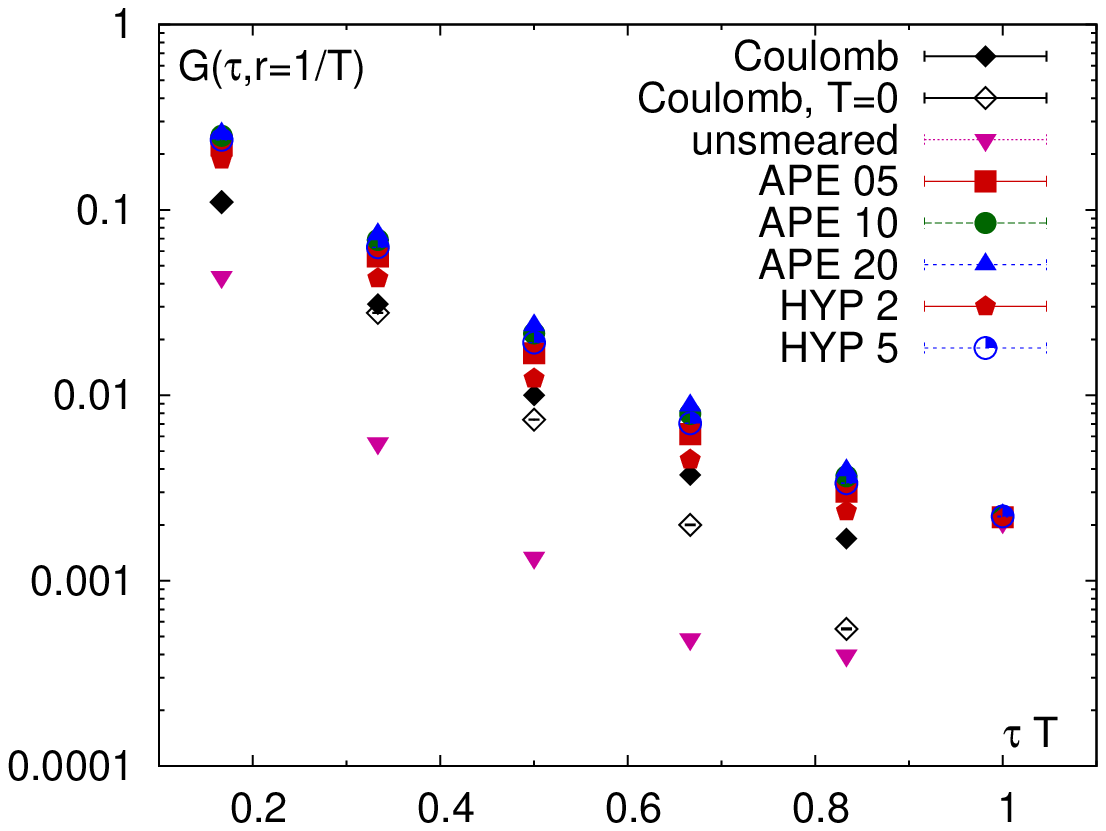}
\end{center}
\caption{Static meson correlators as function of $\tau$ for two spatial separations:
$rT=1/2$ (upper panels) and $r T=1$ (lower panels). The numerical results are shown
for two temperatures $T=147$ MeV (left) and $T=266$ MeV (right).}
\label{fig:wl}
\end{figure*}

\section{Static quark anti-quark correlators in lattice QCD}

Let us consider a static meson operator of the form
\begin{equation}
O(x,y;\tau)=\psi^{\dagger}(x,\tau)U(x,y;\tau) \psi(y,\tau).
\label{meson}
\end{equation}
Here $\psi$ and $\psi^{\dagger}$ are the static quark and 
anti-quark fields, and $U(x,y;\tau)$ is the spatial Wilson
line connecting points $x$ and $y$. After integrating
out the static quark anti-quark fields the 
correlation function of this operator at time $\tau$
becomes the well known Wilson loop of size $r \times \tau$ (see e.g. Ref. \cite{Jahn:2004qr}).
One can also
fix the Coulomb gauge and omit the spatial Wilson
line $U(x,y;\tau)$. In this case one gets correlation
function of two Wilson lines of extent $\tau$.
The static meson correlation function 
has the following spectral representation \cite{Jahn:2004qr}
\begin{equation}
G(r,\tau)=\sum_{n=1}^{\infty} c_n(r) \exp(-E_n(r) \tau),~~r=|x-y|.
\label{spectral0}
\end{equation}
The coefficients $c_n$  depend on the choice of the operator, while
the energy levels $E_n(r)$ corresponding to all possible states containing
static quark and anti-quark do not \cite{Jahn:2004qr}.
If truncated to include some number of terms, the above spectral representation can  in principle 
be used to determine
the energy levels of $Q\bar Q$ states in QCD at zero temperature.
The lowest energy level of $Q\bar Q$ pair is commonly referred to as the static
potential, while the higher energy levels are called the hybrid potentials. 
At sufficiently high temperatures, however, Eq. (\ref{spectral0}) is of little use since 
$\tau \le 1/T$ and one has to consider infinite number of terms in
Eq. (\ref{spectral0}). As mentioned above at finite temperature the energy of 
a static $Q \bar Q$ pair has also an imaginary part. It is not clear how
this imaginary part would manifest itself in the Euclidean correlation function.
It was suggested to generalize the spectral
representation of the static meson correlators in terms of a temperature 
dependent spectral function \cite{Rothkopf:2009pk,Rothkopf:2011db}
\begin{equation}
G(r,\tau)=\int_{-\infty}^{\infty} d \omega \sigma_r(\omega,T) e^{-\omega t}.
\label{spectral}
\end{equation}
At zero temperature Eq. (\ref{spectral0}) and Eq. (\ref{spectral}) are equivalent
and the spectral function $\sigma(\omega,T)$ is just the sum of delta functions.
The imaginary part of the static energy now is encoded in the width of
the spectral function. While the spectral function $\sigma_r(\omega,T)$ depends
on the choice of the meson operator $O(x,y;\tau)$, the peak position and its
width will not as long as there is a well defined bound state peak.
Another way to arrive at Eq. (\ref{spectral}) is to consider the correlation
function of static meson operators in real time 
\begin{eqnarray}
&
\displaystyle
D^{>}(t)={\rm Tr}( O(t) O(0) \exp(-\beta H))=\nonumber\\
&
\displaystyle
\int_{-\infty}^{\infty} \frac{d \omega}{2 \pi} 
\exp(-i \omega t) D^{>}(\omega)
\end{eqnarray}
and then continue to imaginary time $D^{>}(t) \rightarrow D^>(-i\tau)$. Using the standard definition
of the spectral function
\begin{equation}
\sigma_r(\omega,T)=\frac{D^{>}(\omega)-D^{<}(\omega)}{2 \pi}
\end{equation} 
and recalling that $D^{<}=0$ in the case of static quarks (see discussions in Ref. \cite{Brambilla:2008cx})
we get Eq. (\ref{spectral}). Note, that unlike quarkonium spectral function $\sigma_r(\omega)$ is not
odd function of $\omega$.
From the point of view of MEM reconstruction of the spectral function $\sigma_r(\omega,T)$
from Eq. (\ref{spectral}) is not much different from the reconstruction of the quarkonium spectral functions.
However, the structure of the spectral function in the static case could be simpler
and the cancellation of the temperature effects in the
correlation function due to subtle interplay  between the bound state peak
and the continuum is not expected to happen. As a result
we should see a more pronounced temperature dependence of the correlation functions.
The above discussion holds for $\tau<1/T$. The case $\tau=1/T$ 
should be considered separately because it is related to the free energy of static $Q\bar Q$ pair.

The correlation function evaluated at the maximal Euclidean time extent
gives the so-called singlet free energy $F_1(r,T)=-T \ln G(r,\tau=1/T)$.
The singlet free energy has been studied in the past in pure gauge theory
\cite{Zantow:2001yf,Kaczmarek:2002mc,Digal:2003jc,Kaczmarek:2003dp,Kaczmarek:2004gv,Bazavov:2008rw} and in QCD
\cite{Petreczky:2004pz,Kaczmarek:2005ui,Kaczmarek:2007pb,Petreczky:2010yn}.
Most of these studies use Coulomb gauge correlators but in Refs. \cite{Zantow:2001yf,Bazavov:2008rw}
also Wilson loops have been considered.
Unlike the true free energy of a static quark anti-quark pair $F(r,T)$,
defined in terms of Polyakov loop correlators, $F_1(r,T)$ depends on
the choice of the gauge or the choice of $U(x,y;\tau)$ and thus is not
physical. However, in the 
limit of the infinite separation it gives the $Q \bar Q$ free energy
$F_{\infty}(T)=\lim_{r \rightarrow \infty} F(r,T)= \lim_{r \rightarrow \infty} F_1(r,T)$.
The singlet free energy is 
gauge independent in hard thermal loop (HTL) approximation \cite{Petreczky:2005bd}
and coincides with the real part of the static energy in this approximation 
\cite{Laine:2006ns,Brambilla:2008cx}.
The singlet free energy also arises in the context of short distance, $r T\ll 1$
behavior of the Polyalov loop correlator and its decomposition into
singlet and octet contribution within the pNRQCD framework \cite{Brambilla:2010xn}.
Here it is defined as the singlet field correlator and thus is independent of $U(x,y;\tau)$
or gauge fixing. It is equal to the zero temperature static potential up to power corrections,
which are well defined and calculable at any order of perturbation theory. 
The singlet free energy defined in Coulomb gauge or in terms of rectangular (unsmeared) Wilson 
loops was studied using resummed HTL perturbation theory \cite{Burnier:2009bk}. 
This calculation revealed
that new type of temperature dependent divergences, so-called intersection divergences
appear in the case of Wilson loops.
These divergences have been shown to arise due to mixing of cyclic Wilson loops with
the Polyakov loop correlators \cite{Berwein:2012mw} and can be removed if a proper renormalization procedure
is implemented. This should make a quantitative comparison of the perturbative results 
with lattice data on the singlet free energy possible, thus providing
a stringent test of the weakly interacting nature
of QGP.

\section{Numerical results}
\subsection{Lattice approach and parameters}
We calculated static meson correlation functions on $24^3 \times 6$ lattice.
Because of relatively small $N_{\tau}$ small statistical errors can be achieved.
The gauge configurations used in our study have been generated by the HotQCD collaboration
using a combination of the tree-level improved gauge action and Highly Improved Staggered Quark (HISQ) action 
\cite{Bazavov:2011nk}. This combination of the gauge action and quark action was referred
to as the HISQ/tree action in Ref. \cite{Bazavov:2011nk} but here we refer to it
as the HISQ action for simplicity. The Highly Improved Staggered Quark action was first discussed
in Ref. \cite{Follana:2006rc}. The gauge configurations have been generated using 
rational hybrid Monte-Carlo algorithm \cite{Clark:2004cp}.
The algorithmic details of dynamical HISQ simulations can be found in Ref. \cite{Bazavov:2010ru}.
The simulations  have been performed for the physical value of
the strange quark mass $m_s$ and light quark masses $m_l=m_s/20$. This light quark
mass corresponds to the pion mass of $160$ MeV in the continuum limit \cite{Bazavov:2011nk}.
The parameters of the lattice simulations including the lattice gauge coupling $\beta=10/g^2$
and the strange quark mass in lattice units are shown in  Table \ref{tab1} along with
the corresponding temperatures. 
As one can see from the table we consider a wide temperature range across
the chiral crossover, which for $N_{\tau}=6$ occurs at $T_c=171(1)$ MeV \cite{Bazavov:2011nk}.
The last column of the table shows the accumulated statistics
for each $\beta$ value in terms of molecular dynamics time units (TU). 
Static meson correlators have been calculated after each 10 TUs.
The lattice spacing $a$ is determined
from the $r_1$ parameter defined in terms of the zero-temperature static potential as
\begin{equation}
\left.r^2 \frac{d V}{d r}\right|_{r=r_1}=1.0,
\end{equation}
and we use the value $r_1=0.3106$~fm \cite{Bazavov:2010hj}.
We use the  parameterization of the lattice spacing and the quark masses as functions of the gauge  coupling 
$\beta$ along the lines of constant physics that are
given in Ref. \cite{Bazavov:2011nk}. 
\begin{table}
\caption{Parameters of the lattice calculations}
\label{tab1} 
\begin{center}      
\begin{tabular}{llll}
\hline\noalign{\smallskip}
$\beta$ & $m_s$  &  $T$ [MeV] & \# TU   \\
\noalign{\smallskip}\hline\noalign{\smallskip}
6.000   & 0.1138 &  147       &  27000 \\
6.050   & 0.1064 &  155       &  30000 \\
6.100   & 0.0998 &  162       &  30000 \\
6.150   & 0.0936 &  170       &  30000 \\
6.195   & 0.0880 &  178       &  30000 \\
6.285   & 0.0790 &  194       &  30000 \\
6.423   & 0.0670 &  223       &  30000 \\
6.575   & 0.0564 &  258       &  30000 \\
6.608   & 0.0542 &  266       &  30000 \\
6.664   & 0.0514 &  281       &  30000 \\
6.800   & 0.0448 &  320       &  7000 \\
6.950   & 0.0386 &  368       &  7480 \\
7.150   & 0.0320 &  442       &  4770 \\
7.280   & 0.0284 &  488       &  4310 \\
\noalign{\smallskip}\hline
\end{tabular}
\end{center}
\end{table}

Typically in lattice calculations a straight line path is used in the definition of
the static meson operator in Eq. (\ref{meson}), i.e. rectangular Wilson loops being
calculated on the lattice. The main problem with calculating rectangular Wilson loops
on the lattice and extracting potential from them is the large noise. To reduce
the noise and improve the signal for the ground state potential smeared gauge links
are used. The simplest example of smeared gauge links is the well known APE smearing \cite{Albanese:1987ds}.
In this case a 3-link staple is added to each elementary lattice gauge link with some coefficient
and the resulting sum is projected to SU(3) gauge group. Another smearing called hyper-cubic or HYP
smearing was introduced in Ref. \cite{Hasenfratz:2001hp}. In this case the smeared links are constructed 
from APE smeared links in lower dimensional subspace, such that the entire construction of the smeared
links stays within the hypercube \cite{Hasenfratz:2001hp}. This smearing procedure depends on 3 parameters in 4 dimensions,
one for each level of APE smearing. Both APE and HYP smearing can be applied iteratively to increase
the signal to noise ratio. Typically several steps of APE and HYP smearing are used. 
The HYP smearing turned out to be more efficient than the APE smearing in the calculations of the
static potential in pure gauge theory \cite{Hasenfratz:2001hp}. The APE smearing was used in
calculating the static potential in 2+1 flavor QCD (see e.g. \cite{Cheng:2007jq,Cheng:2009zi}),
while HYP smearing was used for example in Refs. \cite{Detmold:2008bw}. 
An alternative approach to reduce the noise in the calculations of the static potential is
to fix the Coulomb gauge (see e.g. \cite{Bazavov:2011nk,Aubin:2004wf}). Both approaches
turned out to be equally efficient.

In the present study we calculated static meson correlators $G(\tau,r)$ at non-zero temperature 
using smeared spatial gauge links as well as Coulomb gauge. 
The smearing of the gauge links was restricted to spatial directions only, i.e. no
temporal links enter the construction of the smeared links. 
We used several iterations of 
APE and HYP smearing.  
The coefficient of the staple for the APE
smearing was $c=0.4$, while for the HYP smearing we used 
the same smearing parameters as in Ref. \cite{Hasenfratz:2001hp}, i.e. $\alpha_2=0.3$ and $\alpha_3=0.6$
in the notation of Ref. \cite{Hasenfratz:2001hp}. The parameter $\alpha_1$ does not enter because we
consider spatial directions only. 
We used $5,~10$ and $20$ iterations  of APE smearing and $1,~2$ and $5$ iterations of HYP smearing.  
\begin{figure}
\includegraphics[width=7.5cm]{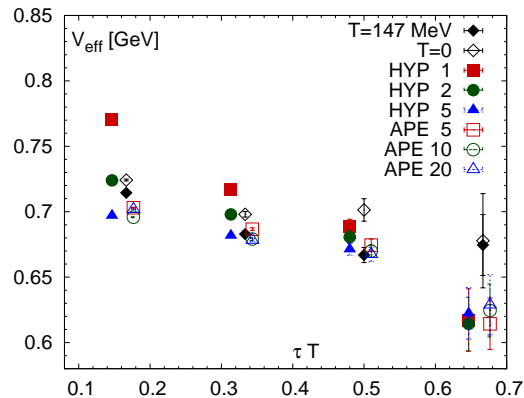}
\caption{The effective potential for $r T=1/2$ as function of $\tau$
at the lowest temperature $T=147$ MeV. The open and filled diamonds correspond to
Coulomb gauge correlators at zero and non-zero temperature, respectively.}
\label{fig:veff_low}
\end{figure}
\begin{figure*}
\begin{center}
\includegraphics[width=7.5cm]{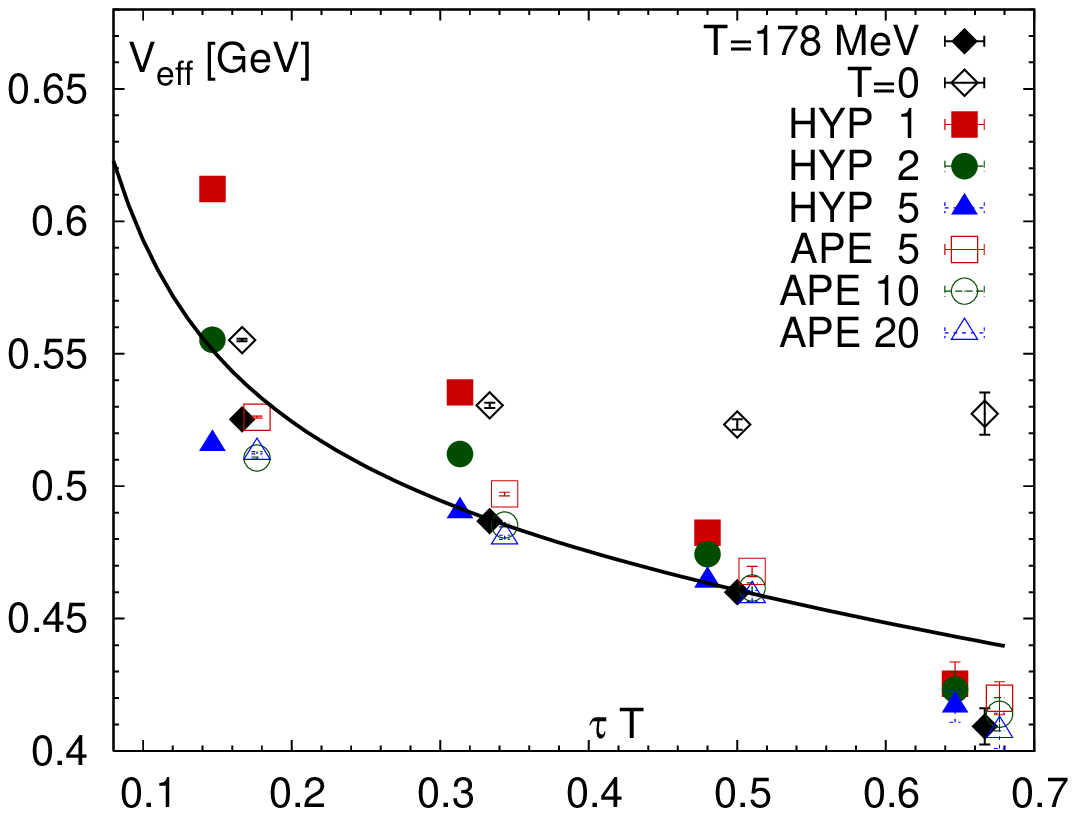}
\includegraphics[width=7.5cm]{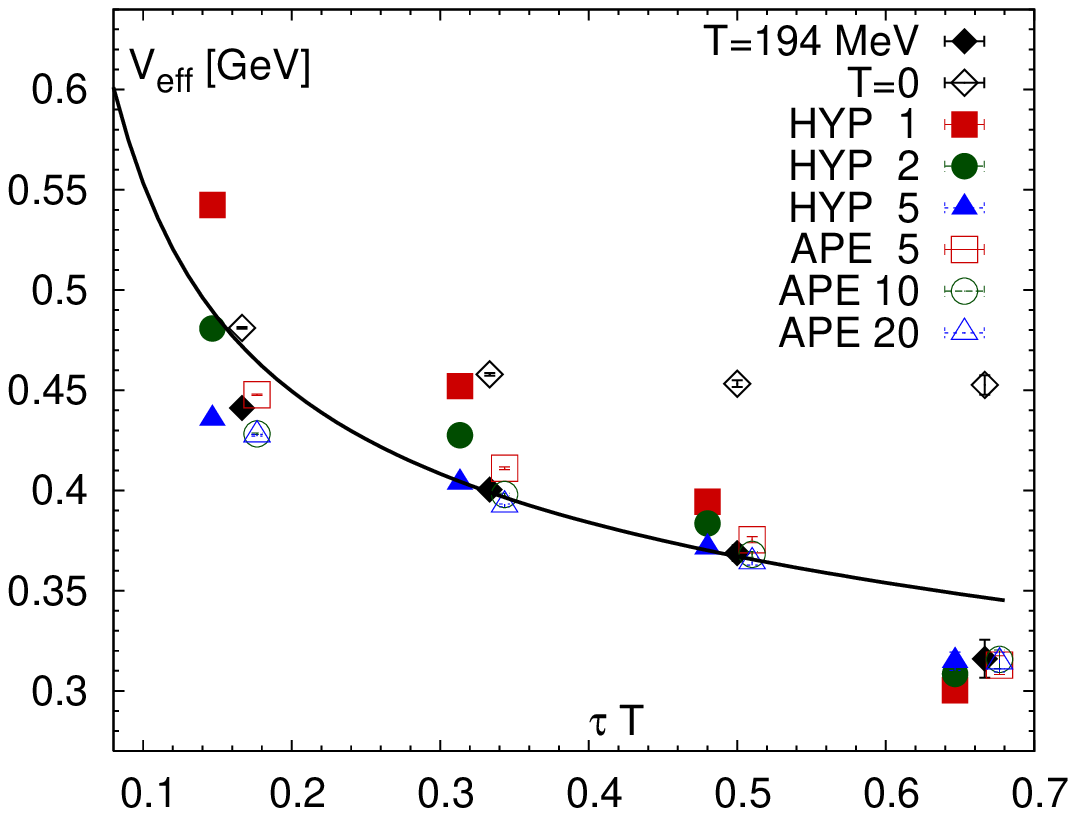}
\end{center}
\begin{center}
\includegraphics[width=7.5cm]{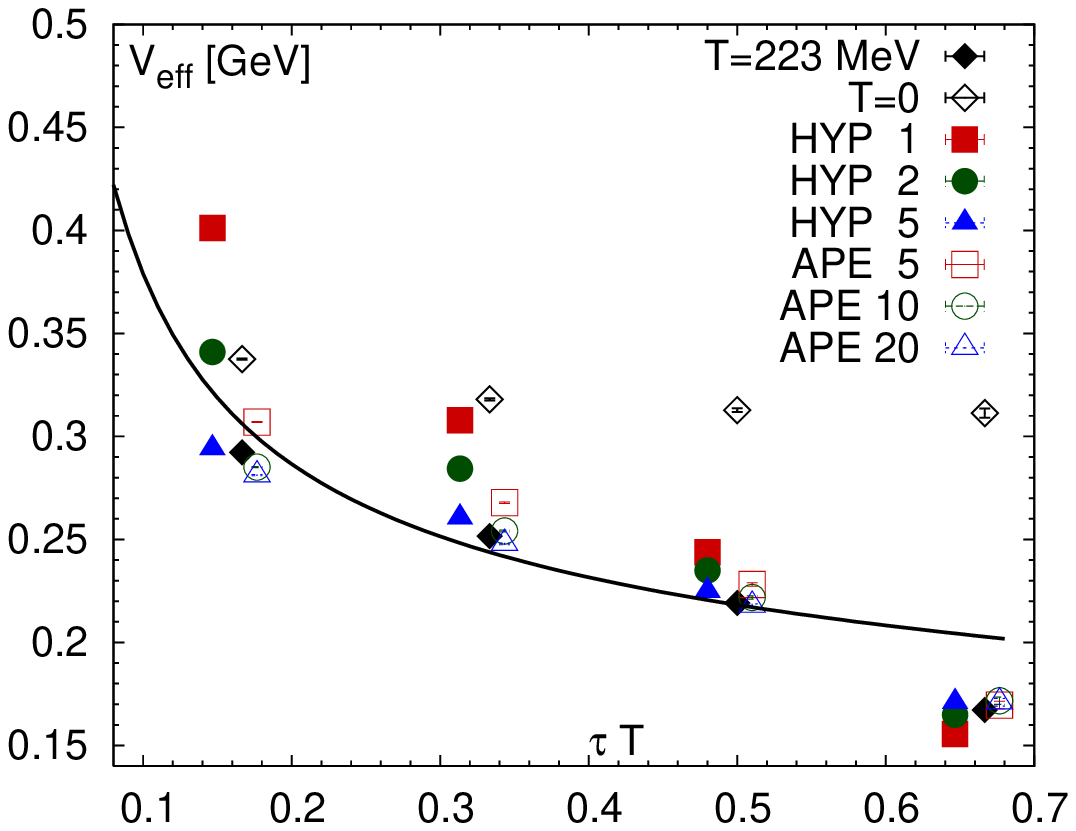}
\includegraphics[width=7.5cm]{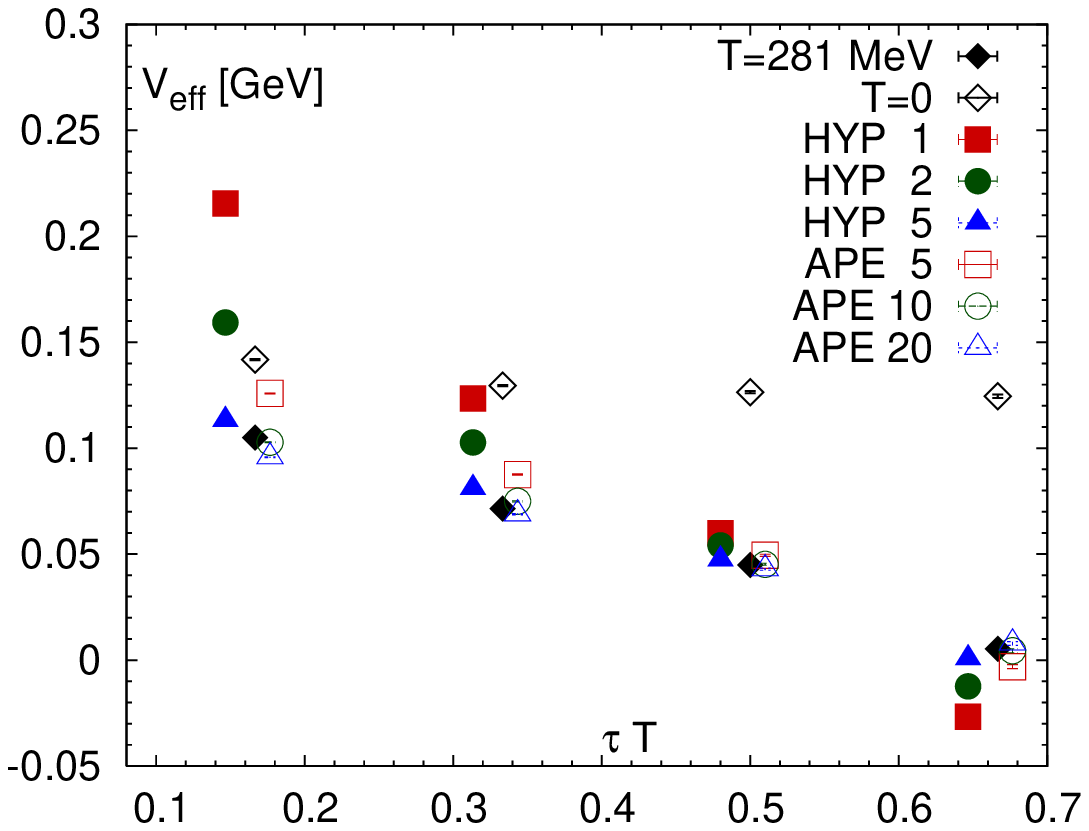}
\end{center}
\caption{The effective potential corresponding to various correlators 
at different temperatures for $r T=1/2$ as function of $\tau$.
The open and filled diamonds correspond to
Coulomb gauge correlators at zero and non-zero temperature, respectively.
}
\label{fig:veff_9}
\end{figure*}
\begin{figure*}
\begin{center}
\includegraphics[width=7.5cm]{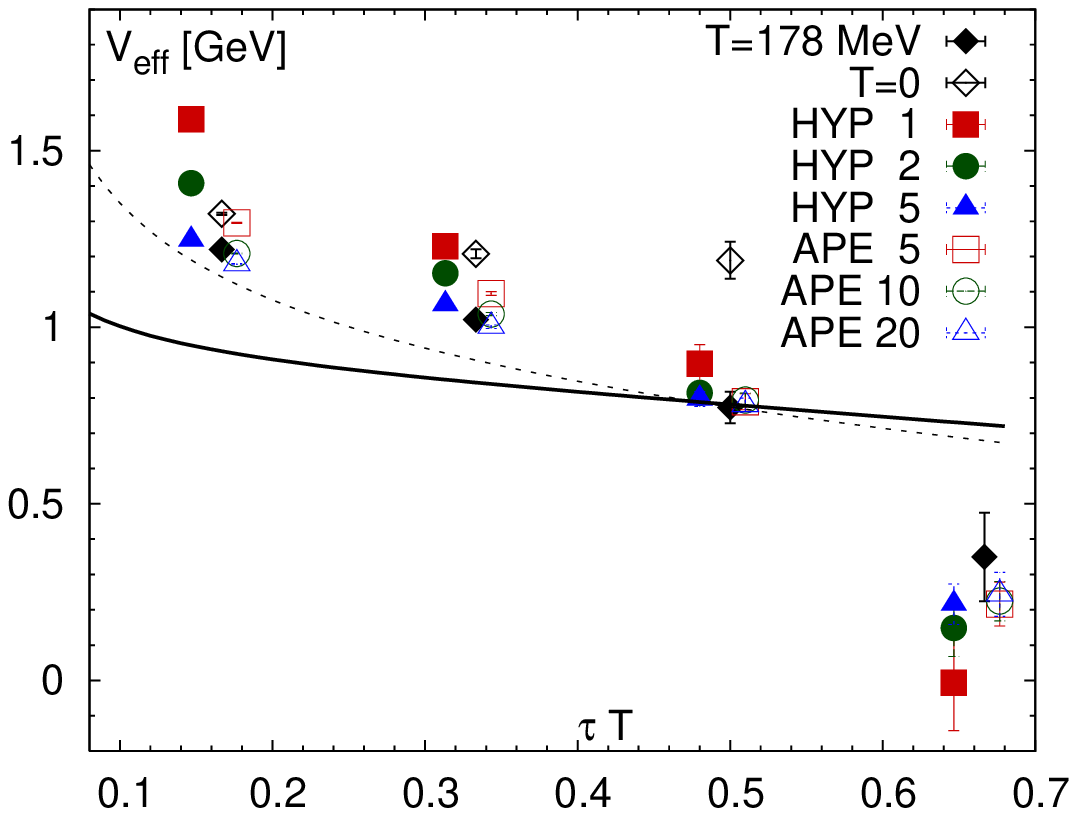}
\includegraphics[width=7.5cm]{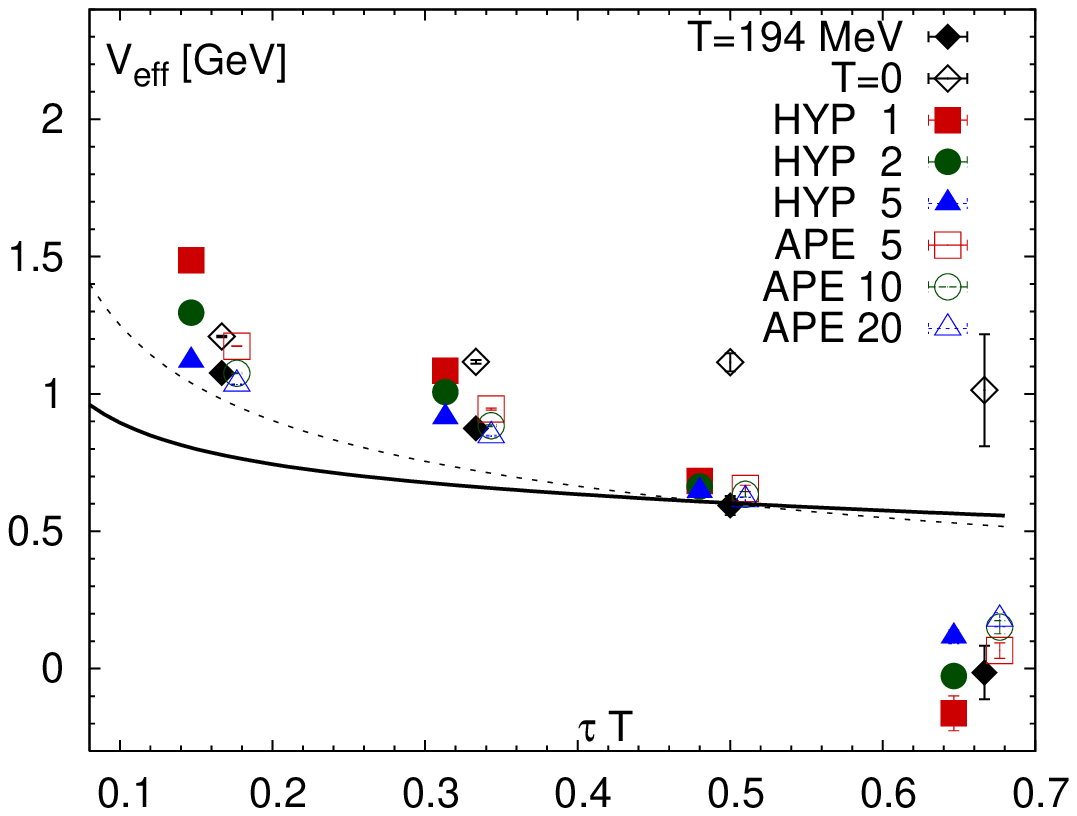}
\end{center}
\begin{center}
\includegraphics[width=7.5cm]{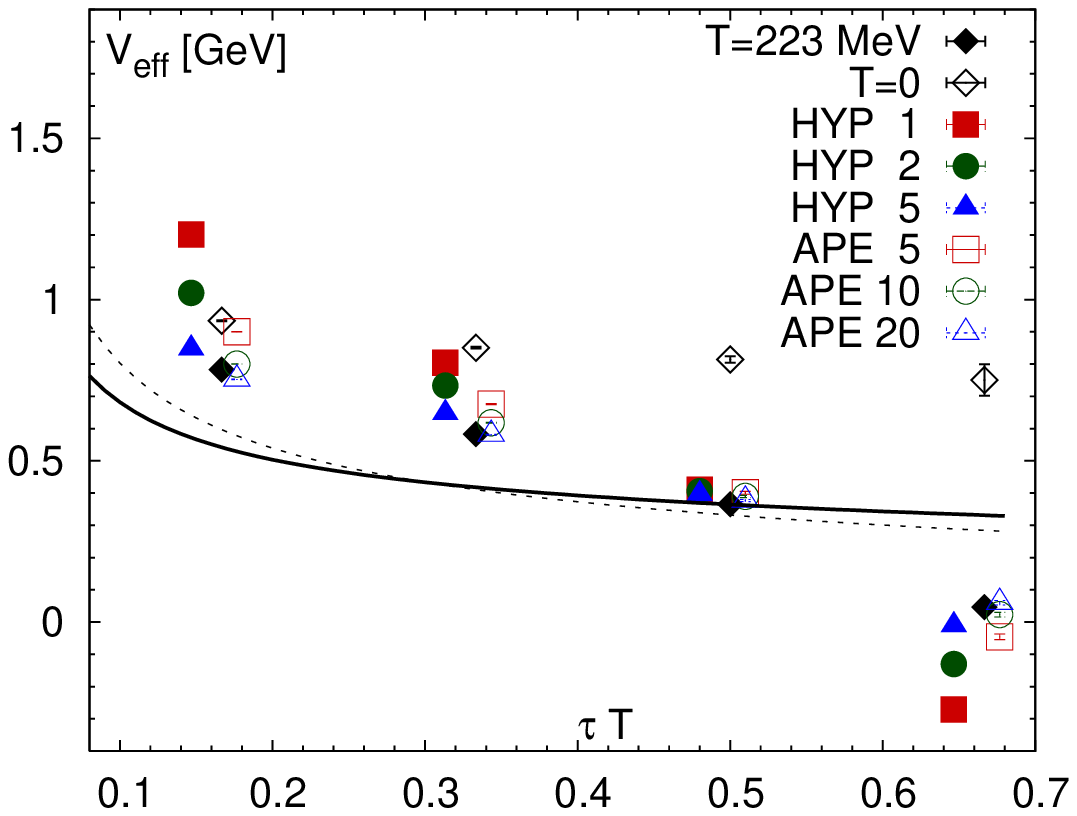}
\includegraphics[width=7.5cm]{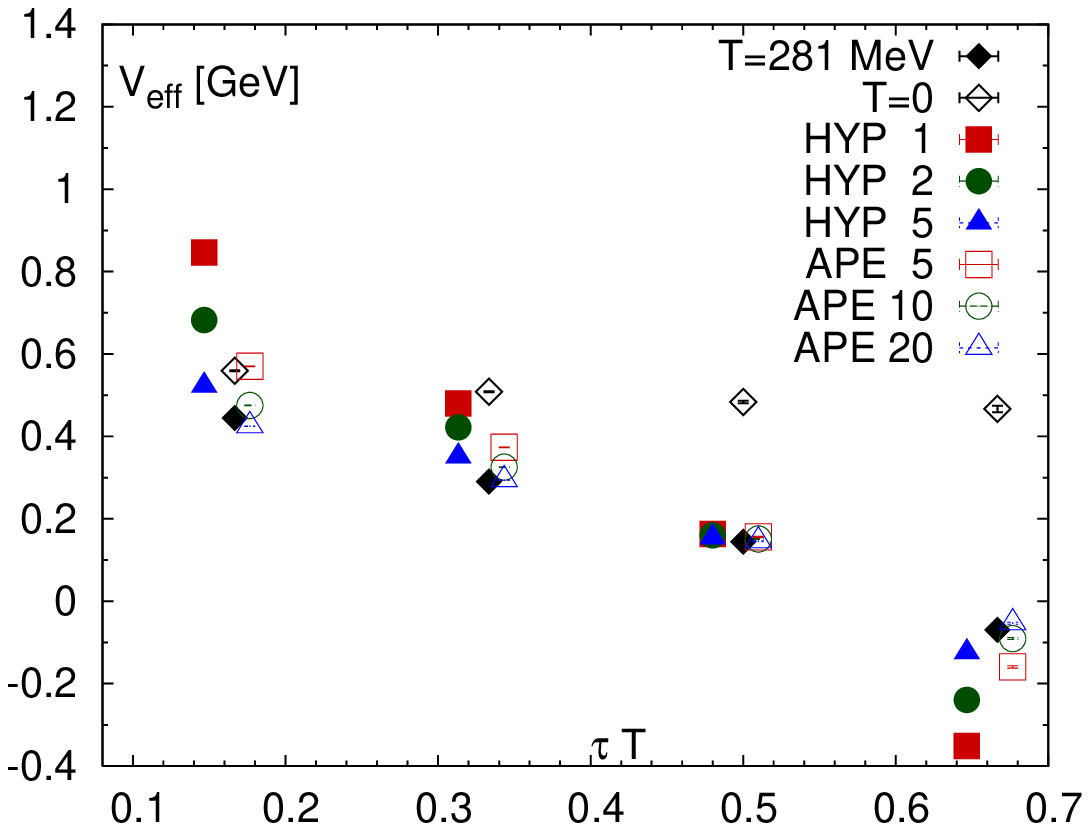}
\end{center}
\caption{The effective potential corresponding to various correlators
at different temperatures for $r T=1$ as function of $\tau$.
The open and filled diamonds correspond to
Coulomb gauge correlators at zero and non-zero temperature, respectively.
}
\label{fig:veff_36}
\end{figure*}

\subsection{Time dependence of static correlatots}
As the first step toward understanding the temperature dependence of the static energy
one should examine the $\tau$-dependence of static meson correlators for various
temperatures and different spatial separations  $r$. As the medium effects on static
meson correlators are
expected to depend on how $r$ and $\tau$ compare to the inverse temperature we will scale these variables
by the temperature when presenting our numerical results.  
Our numerical results for the static meson correlators as function of $\tau$ 
with different smeared links as well as for Coulomb gauge are shown in Fig. \ref{fig:wl}
for two spatial separation $r T=1/2$ and $r T=1$.
We also compared our results with zero temperature results obtained in Coulomb gauge \cite{Bazavov:2011nk}. 
Note that unlike quarkonium correlators static meson correlators are not subject to 
periodic boundary conditions and thus can be studied also for $\tau T>1/2$. 
As expected we see significant in-medium modification of the static meson correlators in contrast to
quarkonium correlators. The modifications become larger with increasing $r$
and increasing temperature: they are negligible at $rT=1/2$, $T=147$ MeV and are the largest
for $rT=1$ and $T=266$ MeV. While the correlators in Coulomb gauge and smeared Wilson loops 
appear to be similar, unsmeared Wilson loops show quite different behavior. They appear to
have more curvature as function of $\tau$ and also have a different slope. 
One of the most prominent features of the finite temperature data is  that  
the static meson correlator does not seem to follow an exponential decay
at large $\tau$ but flattens off or slightly increase around $\tau T \simeq 1$. 
\begin{figure}
\begin{center}
\includegraphics[width=7.5cm]{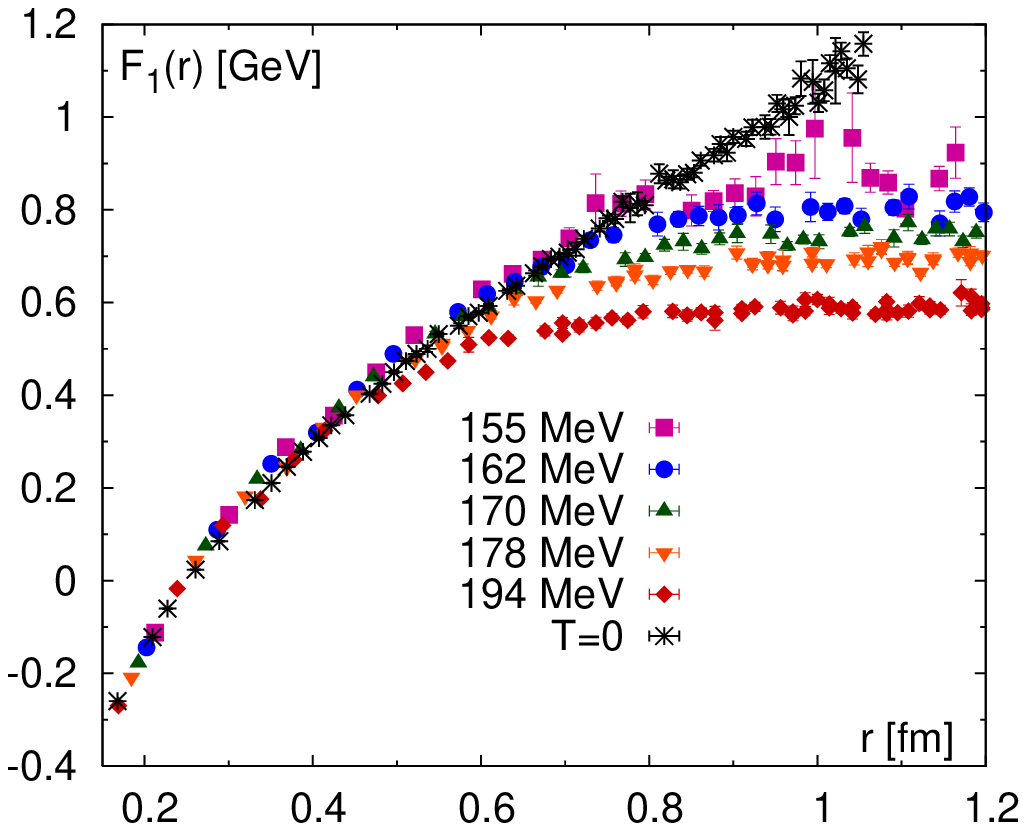}
\includegraphics[width=7.5cm]{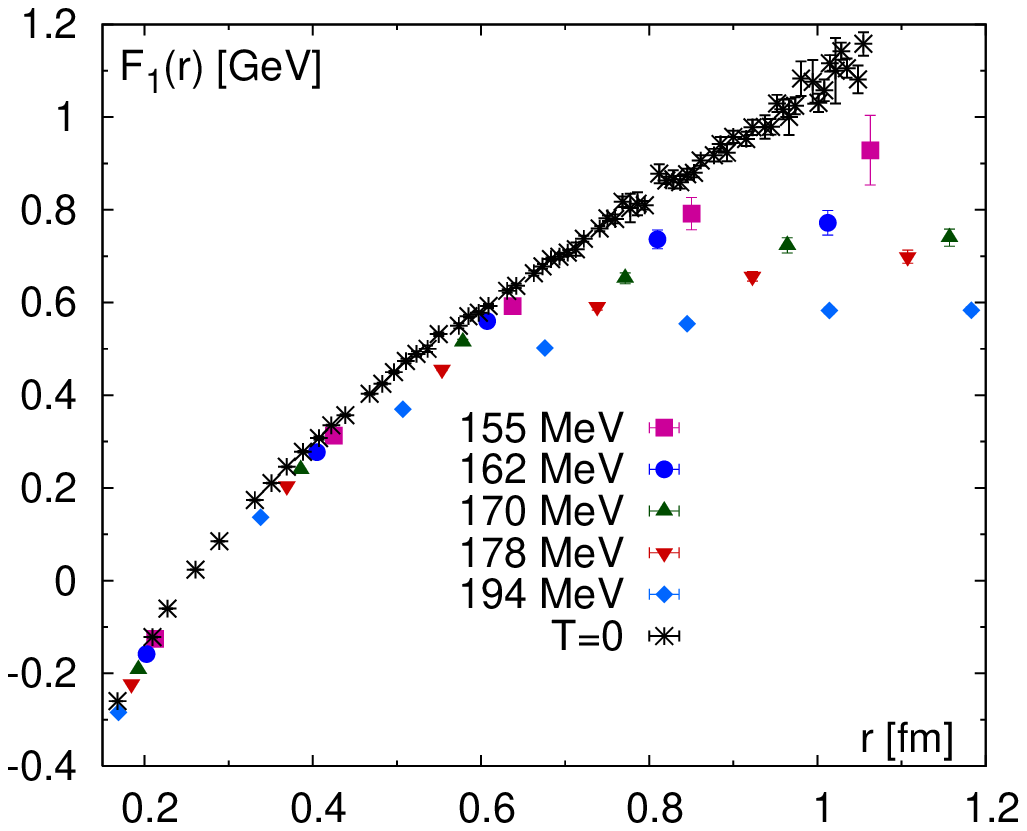}
\end{center}
\caption{The singlet free energy as function of distance $r$ for different temperatures
$T<200$ MeV calculated in Coulomb gauge (top) and using smeared Wilson loops with 5 steps of HYP smearing (bottom).
We also show the zero temperature static potential for $\beta=6.664$\cite{Bazavov:2011nk}. }
\label{fig:f1low}
\end{figure}

The increase around $\tau T \simeq 1$ can be interpreted as a small contribution to
the spectral function $\sigma(\omega)$ at negative frequencies which arises due to
periodic boundary conditions on the gluon background even-though static quarks propagate
only forward in time, i.e. correspond to positive frequency. At least such interpretation seems
to be supported by MEM analysis of Wilson loops \cite{Rothkopf:2011db} 
and bottomonium correlators obtained in NRQCD \cite{Aarts:2010ek}. 
Another way to understand the increase of the spatial meson correlator at $\tau T=1$ is
to inspect the HTL resummed perturbative result for rectangular Wilson loop \cite{Laine:2006ns}.
The resummed 1-loop correction to the rectangular Wilson loop consist of three terms:
a term linear in $\tau$ that is proportional to the real part of the potential, a constant
term and a term periodic in $\tau$ \cite{Laine:2006ns}.
After a proper re-exponentiation the term linear in $\tau$
will correspond to the exponentially decaying part, while the two other terms can explain
the behavior at $\tau T \simeq 1$. 

While different correlators have different behavior in $\tau$ they all converge at a similar
value for $\tau T=1$. 
For very large distances in fact this is expected, as $-T \ln G(\tau=1/T,r)$ gives the 
free energy of static quark anti-quark pair and therefore all correlators should be equal
(see the discussion in the previous section).
The convergence of different correlators for $r T=1$ is due to strong screening effects, i.e.
the fact that $-T \ln G(\tau=1/T,r)$ is not very different from $F_{\infty}$.

At zero temperature the large $\tau$ behavior of $G(r,\tau$) is dominated by the static potential.
Therefore it is customary to study the effective potential defined in terms of the following
ratio
\begin{equation}
a V_{eff}(r,\tau)=\ln \frac{G(r,\tau/a)}{G(r,\tau/a+1)}.
\end{equation} 
Here $a$ denotes the lattice spacing. The effective potential is expected to decrease
with increasing $\tau$, eventually reaching a plateau that corresponds to the value of
the static potential. How quickly $V_{eff}$ reaches a plateau depends on the choice
of the static meson operator in Eq. (\ref{meson}), using more smearing in the gauge
connection $U(x,y;\tau)$ typically suppresses the contribution from the excited states.
Using the Coulomb gauge is also an effective method for suppressing the excited state
contributions. 
The effective potential turns out to be  a useful tool to examine the behavior of the
static meson correlators in more detail. 
In Fig. \ref{fig:veff_low} we show our numerical results 
for $V_{eff}$ calculated for different smearing of the spatial links as well as 
for Coulomb gauge for the lowest
temperature of $147$ MeV and distance $r T=1/2$. 
We do not show $V_{eff}$ from unsmeared Wilson loops as the corresponding numerical
results are too noisy. We compare our numerical results with the zero temperature
results in Coulomb gauge \cite{Bazavov:2011nk} which approach a plateau within the
considered $\tau$ range. At the lowest temperature we do not expect
large modification of the static potential. Indeed the temperature effects on $V_{eff}$ are
small for $\tau T<1/2$ and indicate a downward shift of the static potential of about $10$ MeV.
This is  consistent with the study of the static potential in pion matter, where a similar
downward shift has been observed \cite{Detmold:2008bw} (see also Ref. \cite{Detmold:2012pi}
for a related study for quarkonium). 
The figure also shows that Coulomb gauge results and smeared Wilson loops give similar
results if sufficient number of smearings steps is used, i.e. 5 or more smearing steps
in the case of APE smearings and more than two steps for HYP smearings. 
For one or two steps of HYP smearings the excited state contribution, seen
as the curvature at small $\tau$,  is significant.
While the effective potential reaches a plateau at zero temperature within the studied
$\tau$-range, this is not the case for the lowest temperature. 
The decrease in $V_{eff}$ at $\tau T=2/3$ is particularly striking and difficult
to explain in terms of the expected in-medium modifications of the static energy.
Similar results have been obtained for $T=155$ MeV, $162$ MeV and $178$ MeV.

Now let us discuss the behavior of $V_{eff}$ above the chiral transition temperature.
We show our numerical results for $r T=1/2$ and $r T=1$ in Fig. \ref{fig:veff_9}
and Fig. \ref{fig:veff_36}, respectively. The features of $V_{eff}$ discussed above
also hold at higher temperatures, in particular, $V_{eff}$ does not show a plateau 
and there is a rapid decrease around $\tau T=2/3$. The effective potential above $T_c$ 
is always smaller than the $T=0$ result and the deviations increase
with increasing temperatures and distance suggesting that the static potential decreases
with increasing temperature. 
Note that we do not present our results for $V_{eff}$ from unsmeared Wilson loops as these show
much stronger $\tau$-dependence and would change the scale of Figs. \ref{fig:veff_9}
and \ref{fig:veff_36} substantially.

At temperatures significantly higher than the transition temperature we expect 
the imaginary part of the potential to play an important role.
Therefore it is natural to ask whether the above features of $V_{eff}$ are due
to the imaginary part. 
According to thermal pNRQCD the spectral function defined in terms of $D^{>}$, i.e.
$\sigma_r=D^{>}(\omega)/(2 \pi)$  has the Breit-Wigner form \cite{Brambilla:2008cx}
\begin{equation}
\sigma_r(\omega)=\frac{1}{\pi} \frac{{\rm Im} V}{(\omega-{\rm Re} V)^2+{\rm Im }V^2}
\label{BW}
\end{equation}
(c.f. Eq. (96) in Ref. \cite{Brambilla:2008cx}). 
To model the effects of ${\rm Im} V$ on the effective potential we calculated
the static correlators using the above form for the spectral function and Eq. (\ref{spectral})
together with some plausible values of ${\rm Im} V$ inspired by weak coupling calculations.
The $r$-dependence of the imaginary part of the potential is quite complicated even
in the weak coupling regime. For large $r$, however, it has a simple form $C_F \alpha_s m_D$, while
it vanishes for $r T \ll 1$. For the relevant temperature range $\alpha_s m_D \sim 1$. Therefore
in our modeling efforts we assume
that ${\rm Im} V=T$ for $rT=1$ and ${\rm Im} V=T/2$ for $rT=1/2$. The value of
${\rm Re} V$ was adjusted to match the value of $V_{eff}$ at $\tau T=1/2$. The corresponding
results are shown in Figs. \ref{fig:veff_9} and  \ref{fig:veff_36} as solid black lines.
For $r T=1$ we also allow for a larger value of the imaginary part of the potential, namely
${\rm Im} V=3 T$. The corresponding results are shown as the dashed lines 
in Figs. \ref{fig:veff_9} and  \ref{fig:veff_36}. 
As one can see from the figures the $\tau$ dependence of $V_{eff}$ in general is 
not described well by the above
form of the spectral function. For $r T=1/2$ the Breit-Wigner form captures some
features of the $\tau$ dependence but for $rT=1$ the observed curvature of $V_{eff}$ is
much larger than the one that can be provided by the Breit-Wigner form, even if we 
assume that ${\rm Im} V=3 T$. 
The Breit-Wigner form describes the spectral function  well in the vicinity of the pole.
For large values of ${\rm Im} V$ the Breit-Wigner form has long tails that extend to
the $\omega$ regions where the form is not valid, e.g. Eq. (\ref{BW}) obtained from
pNQRCD is not valid for large values of $\omega$.
Therefore we also consider a
Gaussian Ansatz with the same width as used in the Breit-Wigner form and calculate $G(r,\tau)$.
In this case we get very small $\tau$-dependence of $V_{eff}$. Thus the $\tau$ dependence
seen in Figs. \ref{fig:veff_9} and  \ref{fig:veff_36} for the Breit-Wigner form is generated
by its tails. Therefore it seems unlikely that the $\tau$-dependence of $V_{eff}$ is due
to the imaginary part of the static potential, instead there could be an additional contribution
to the static correlator that arises from the periodic boundary.

As discussed above, based on HTL resummed results we expect that Wilson loops
at high temperatures will consist of two contributions: one exponentially
decaying in $\tau$ and another one with more complex $\tau$ behavior.
Therefore we tried to fit the
$\tau$ dependence of the static correlators by a simple form 
\begin{equation}
A \exp(-\tilde V \tau)+c \exp(-\Delta (1/T-\tau))
\end{equation}
for $1/2 \le \tau \le 2/3$.
The fit form works well but  the extracted value of $\tilde V$ depends
somewhat on the chosen fit range. However, it is always smaller than the $T=0$
static potential. This together 
with the analysis of $V_{eff}$ suggests that the energy of static $Q \bar Q$ pair is
decreasing with increasing temperature contrary to the results obtained in
Ref. \cite{Rothkopf:2011db} based on calculations in SU(3) gauge theory and
MEM. We also find that the non-exponential contribution is relatively
small. Namely we find that
$c$ is about 1000 times smaller than $A$, however, it increases 
with increasing temperature.

\subsection{Singlet free energy}

As discussed in section 2  $F_1(r,T)=-T \ln G(r,T)$ gives the singlet free energy of static $Q \bar Q$.
We have calculated the singlet free energy $F_1$ using Coulomb gauge and smeared as well as
unsmeared Wilson loops. We start the discussion of our numerical results first considering
the low temperature region $T<200$ MeV. This region of course also contains temperature values
above the chiral crossover.
Our results for Coulomb gauge and smeared Wilson loops are shown in 
Fig. \ref{fig:f1low} and compared to the zero temperature static energy calculated in
Ref. \cite{Bazavov:2011nk} at $\beta=6.664$.
We do not show the results from unsmeared Wilson loops because they are too noisy. It is suffice
to say that they are above the Coulomb gauge result. 
All the smeared Wilson loops give very similar results for $F_1$, i.e. the details
of smearing do not seem to matter. The singlet free energy does not
show a significant change across the chiral crossover and the temperature effects are relatively
small for $r<0.65$ fm. In the case of Coulomb gauge $F_1$ is slightly
larger than the zero temperature static potential at intermediate distances 
$0.3~{\rm fm} < r < 0.8~{\rm fm}$. This is qualitatively similar to the findings
in pure gauge theory \cite{Digal:2003jc,Bazavov:2008rw,Kaczmarek:2003dp}.
As discussed in Ref. \cite{Bazavov:2008rw} this is due to the fact that $c_1<1$. In
the case of smeared Wilson loops $F_1$ is always smaller than the zero temperature
static energy since $c_1$ is expected to be close to unity \cite{Bazavov:2008rw}.

\begin{figure}
\includegraphics[width=7.5cm]{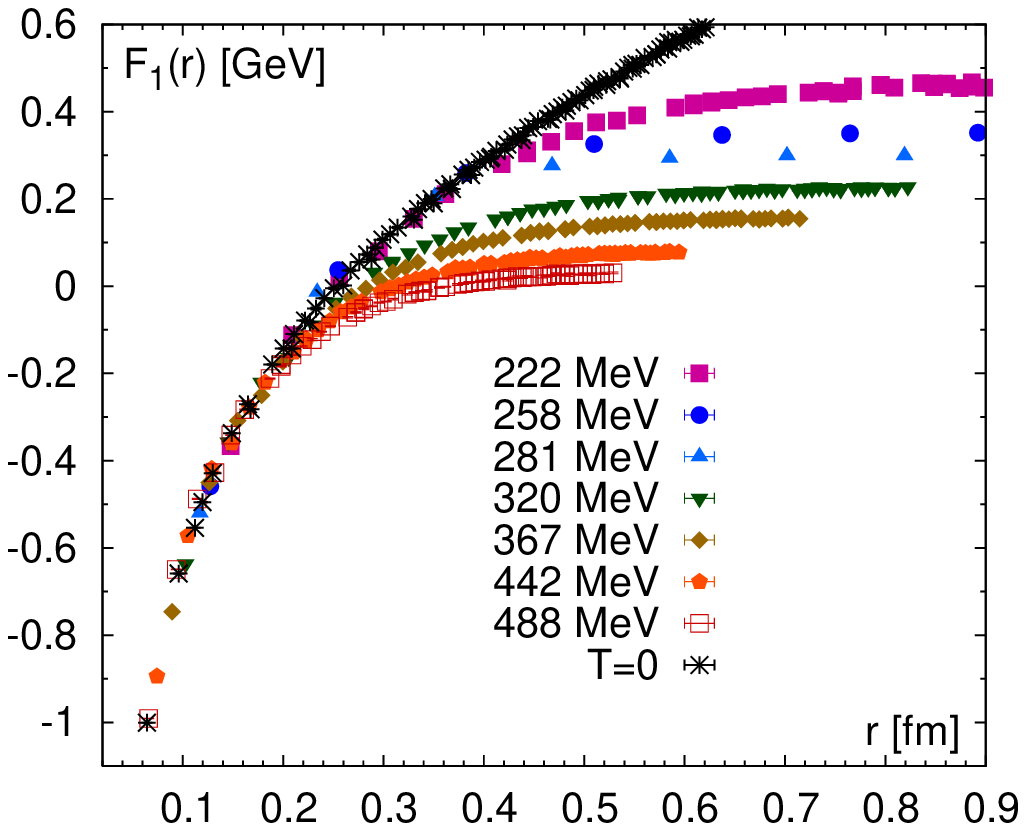}
\includegraphics[width=7.5cm]{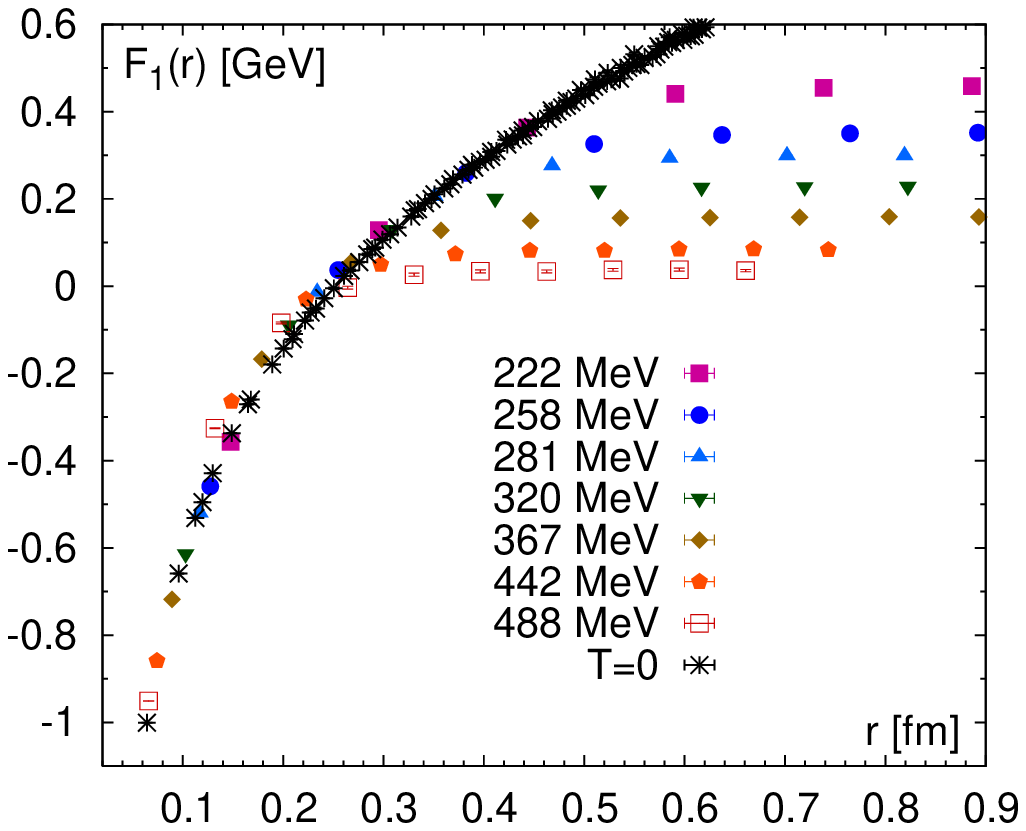}
\caption{The singlet free energy as function of distance $r$ for different temperatures
$T>200$ MeV calculated in Coulomb gauge (top) and using unsmeared Wilson loops (bottom).
We also show the zero temperature static potential for $\beta=6.664$
and $\beta=7.28$ \cite{Bazavov:2011nk}. }
\label{fig:f1high}
\end{figure}

\begin{figure}
\includegraphics[width=7.5cm]{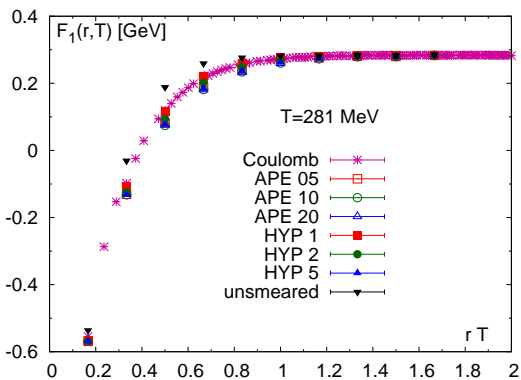}
\caption{The singlet free energy at $T=281$ MeV defined in Coulomb gauge and in terms
of different Wilson loops.}
\label{fig:f1_b6664}
\end{figure}

Our results for the high temperature region, $T>200$ MeV are shown in Fig. \ref{fig:f1high}
for Coulomb gauge and unsmeared Wilson loops. 
As expected at very small distances $F_1$ is temperature independent and coincides with
the $T=0$ potential. For $T>200$ MeV  the numerical
results for unsmeared Wilson loops have quite small statistical errors.
Therefore one can use our numerical results to test the perturbative predictions
for cyclic Wilson loops \cite{Burnier:2009bk,Berwein:2012mw}.
Note that for Coulomb gauge the singlet free energy
does not overshoot the zero temperature static potential 
similarly to the case of pure gauge theory \cite{Digal:2003jc,Kaczmarek:2003dp,Bazavov:2008rw}.
This may imply that the coefficients $c_n$ are not important for the temperature dependence of $F_1$ at high temperatures. 
As the temperature increases the singlet free energy reaches the plateau at
smaller and smaller distances consistent with expectations based on color screening.
Figure \ref{fig:f1high} shows that $F_1$ obtained from unsmeared Wilson loops is systematically
larger than $F_1$ obtained in Coulomb gauge at intermediate distances. This could be due to the intersection divergences of 
cyclic Wilson loops \cite{Berwein:2012mw}.
Smeared Wilson loops on the other
hand give results which are slightly smaller than Coulomb gauge results at intermediate distances,
except for one iteration of HYP smearing. There is no dependence on the smearing if more than
one smearing steps are used. To make
this point clear in Fig. \ref{fig:f1_b6664} we compare $F_1$ obtained for
different correlators at $T=281$ MeV. 
Similar results have been obtained
at other temperatures.  

To explore the large distance behavior of the singlet free energy we consider
the following combination 
\begin{equation}
S_1(r,T)=-r (F_1(r,T)-F_{\infty}(T)). 
\end{equation}
At leading order this combination
should decay exponentially ans its exponential decay is governed by the leading order
Debye mass $m_D^{LO}=g T \sqrt{3/2}$. The perturbative analysis of $F_1$ quantity at next-to-leading order was first
performed in Ref. \cite{Burnier:2009bk} and in the case of Colomb gauge revealed a power-law
behavior. However, as mentioned already the renormalization of the static meson correlators at $\tau=1/T$
is non-trivial. We find that $S_1(r,T)$ decays exponentially for $r>1/T$. The corresponding screening mass
$\tilde m_E/T$ shows only mild temperature dependence for $T>300$ MeV consistent with expectations. Furthermore,
we find that $\tilde m_E/T$ for unsmeared Wilson loops is significantly larger than for Coulomb gauge.
The screening masses obtained from smeared Wilson loops are similar or slightly smaller than the ones
obtained in Coulomb gauge. The only exception is the case when one iteration of HYP smearing is used, where
$m_E/T$ is closer to the the value obtained from unsmeared Wilson loops.

\section{Conclusions}
In this paper we studied various static meson correlators at non-zero temperature.
We analyzed the temperature dependence of these correlators as well as their dependence
on the Euclidean time $\tau$. The $\tau$-dependence of the correlators has been
analyzed in terms of the effective potential $V_{eff}$ for various spatial separations $r$.
For large Euclidean times $V_{eff}$ gives the static potential in the zero temperature case.
We found that static meson correlators defined in Coulomb gauge
and in terms of smeared Wilson loops show  quite similar behavior. Contrary to
the zero temperature case the effective potential does not reach a plateau in the studied
$\tau$ window even at the lowest temperatures. We explored different possibilities
to explain this behavior including the presence of an imaginary part of the static energy.
We conclude that this behavior is due to the additional contribution not related to the static
potential that arises from the periodic boundary condition. To extract the potential
at non-zero temperature we will need to extend this calculation to larger $N_{\tau}$.
We also calculated the singlet free energy and found that its behavior is qualitatively similar to
the previous findings obtained in pure gauge theory. Comparing our numerical results with
perturbative prediction on $F_1$ \cite{Brambilla:2010xn,Burnier:2009bk,Berwein:2012mw} will
be useful to clarify the strongly or weakly interacting nature of the QGP.

\section*{Acknowledgments} 
This work was supported by U.S. Department of Energy under
Contract No. DE-AC02-98CH10886. The numerical simulations
have been performed at NERSC and  on BlueGene/L computers at the New York Center for Computational
Sciences (NYCCS) at Brookhaven National Laboratory. We would like to thank A. Rothkopf
for valuable discussions on the behavior of Wilson loops at $\tau T\simeq 1$.

\bibliographystyle{epjc}
\bibliography{HotQCD}

\end{document}